# Surface charge induced $sp^3 - sp^2$ transformation: new energy optimization universal mechanism of $(4 \times 4)$ reconstruction of stoichiometric GaN(0001) surface.

Paweł Strak[1], Wolfram Miller[2] and Stanisław Krukowski[1]

[1]Institute of High Pressure Physics, Polish Academy of Sciences, Sokołowska 29/37, 01-142 Warsaw, Poland

[2]Leibniz Institute for Crystal Growth (IKZ), Max-Born-Str. 2, D-12489 Berlin, Germany

**Abstract**

*Ab initio* calculations were applied to large size slabs simulations for the determination of the properties of the principal structures of the stoichiometric GaN(0001) surface. The results are different from published previously: stoichiometric GaN(0001) surface structure is characterized by mixed structure in which (3 / 8) top layer Ga atoms remain in standard position with $sp^3$ hybridized bonding while the remaining (5/8) top layer Ga atoms is located in plane of N atoms with $sp^2$ hybridized bonding. This involves charge transfer to the previous one, entailing energy optimization, therefore $Ga - s$ and $Ga - p$ energy difference is driving force for the transition. In large size surfaces Ga atoms create $(4 \times 4)$ reconstruction which is additionally stabilized by strain optimization. Heavy doping in the bulk changes $sp^3$ to $sp^2$ ratio thus confirming the charge transfer mechanism of the reconstruction. The charge transfer energy optimization is universal for all surfaces terminated by $sp^3$ bonded atoms.

Gallium nitride and its solid solutions with InN and AlN continue to remain in the focal point of the research directed towards development of the new generations of the optoelectronic and electronic devices. These optoelectronic devices include: light emitting diodes (LEDs), laser diodes (LDs) and superluminescent emitting diodes (SLEDs) potentially covering the entire visible and a large portion of UV spectral range [1,2]. In parallel, the nitride based electronic devices designed to be developed include: high power and also high electron mobility transistors (HEMTs) [3]. Thus, the nitride based devices create important contribution to the presently emerging new technology civilization based on carbon-free energy production, distribution and consumption [4].

The elaboration of such devices is critically related to fabrication of perfect nitride structures, predominantly by epitaxial growth on metal-polar GaN(0001) or AlN(0001) surfaces. The quality and the properties of such structures are critically dependent of the properties of such surfaces during epitaxy [5]. In addition the properties of these surfaces are intrinsically related to HEMT functioning. Thus the clean or more precisely stoichiometric surfaces are of great interest and therefore they should be investigated. Nevertheless, it is extremely difficult, or even close to impossible, to attain clean metal polar GaN(0001) or AlN(0001) polar surfaces [6].

Accordingly, an alternative, theoretical route was pursued. That was based on *ab initio* calculations of the finite size slabs representing polar nitride surfaces. In the early stage, the available power of the computers limited the size of the slabs used in the simulations to (2 x 2) systems. Thus, the early *ab initio* determined properties of clean GaN(0001) indicated of absence of any reconstruction [7-9], which was later changed to identification of 2 x 1 row structure having slightly lower energy [10,11]. Both configurations are characterized by gallium broken bond large dispersion quantum state, located in the upper part of the energy gap. The state is partially filled pinning the Fermi level at the surface i.e. the surface is metallic. Thus it should be active which is not compatible with the experimental data, according to which the Ga terminated GaN polar surface is chemically resistant [12]. Much later, the possible explanation of the surface inactivity was attributed to full oxygen coverage [13].

In the following we will present *ab initio* determined structures of the bare, i.e. stoichiometric GaN(0001) Ga-terminated surface. The surface structure will be determined using different slab sizes that induced different reconstructions. The charge analysis will be used to elucidate basic relation between the charge and the reconstructions. A natural extension will be the use of bulk doping to determine the connection between the charge in the bulk and the surface reconstruction. The new relation between reconstruction and the charge will be established.

*Ab initio* density functional theory (DFT) calculations were used in the simulations of the properties of the large size GaN slabs representing Ga-polar stoichiometric GaN(0001) surface. These calculations employ the Spanish Initiative for Electronic Simulations with Thousands of Atoms (SIESTA) package [14]. This *ab initio* software solves Kohn-Sham equations determining its eigenfunctions in linear combination of the numeric atomic orbitals, of the finite, predetermined size in radial direction [15]. The *s* and *p* orbitals of gallium and nitrogen atoms are represented by triple zeta functions. In case of gallium, the internal *d* shell electrons are incorporated in the valence electron set and their eigenstates are represented by single zeta functions. The size of the functional basis was reduced by the Troullier-Martins pseudopotentials replacing the electrons belonging to atomic cores [16,17]. The integration over k-space was replaced by the sum over Monkhorst-Pack special points grid ($1 \times 1 \times 1$) [18]. GGA-PBE (PBEJsJrLO) functional with parameters β, μ, and κ fixed by the jellium surface (Js), jellium response (Jr), and Lieb-Oxford bound (LO) criteria, respectively, as described by [19,20].

The Poisson equation is solved by Fast Fourier Transform (FFT) series method. Therefore periodic boundary conditions are imposed in the solution. The interaction between slab copies was cancelled by additional contribution to the potential offsetting the slab dipole [21, 22]. SCF loop termination condition is the level of the difference for any element of the density matrix not higher than $10^{-4}$. The atom relaxation terminates when the forces acting on the atoms were not higher than 0.005 eV/Å. The *ab initio* simulated GaN lattice constants were $a = 3.21$ Å and $c = 5.23$ Å. These values are in the good agreement with the experimental data for GaN: a = 3.189 Å, c = 5.186 Å [23].

Semiconductor surfaces undergo intensive reconstructions. A principal factor in these transformations is the energy optimization. This is related to the fact that direct termination surface entails existence of the quantum high energy states related to absence of the overlap with missing neighbors [24]. In principle these states (called broken bond states) can be emptied by the charge transfer to the bulk states. Inevitably that leads to surface charge separation which is energetically costly, thus this route is not optimal. Typically, the excess charge at the surface is of order of 0.01 elementary charge for single GaN(0001) surface atom [9]. This is much lower than the charge associated with Ga broken bond which is $1/4\,e$, therefore such charge transfer is not expected to be an efficient route to surface energy minimization.

An alternative way to reduce the energy of surface broken bond states is to reshuffle topmost atoms that create state overlap thus lowering their energy. The exemplary case is the low temperature $(7 \times 7)$ reconstruction of silicon Si(111) surface that includes creation of dimers in the top layer [25]. Thus such surface reconstruction depends critically on the atom bonding. The structure is the result of the energetic interplay between electronic and structural contributions. Generally, additional energy related to the strain is compensated by the electron energy decrease. As shown recently, gallium nitride is not typical $sp^3$ bonded semiconductor – its valence band is composed of two separate subbands: higher - composed of $Ga4sp^3 - N2p$ states and lower - composed of $Ga3d - N4s$ states [26,27]. These states play a different role in the bonding. The lower energy subband, composed of the collection of five $Ga3d$ and one $N4s$ orbitals, is isotropically cohesive. In addition, the higher energy subband is composed of directional $Ga4sp^3$ and off-directional $N2p$ orbitatals. Thus the latter contribution is highly directional and is solely responsible for arrangement of Ga and N atoms into tetrahedral wurtzite lattice. Hence, due to these fraction it is relatively easy to switch from the *sp³* to other configurations at the surface.

The obtained results were likely affected by the size of the simulated system. Generally, it is expected that semiconductor surfaces are reconstructed in order to comply the electron counting rule (ECR) [28]. The surface reconstruction and underlying ECR argument is a result of the energy optimization by the motion of the surface topmost atomic layers. The atoms are moving in such way that the resulting change of the quantum states lowers energy of the occupied states and increases of the empty ones. The latter do not contribute to the total system energy while the change of the occupied ones leads to the total energy decrease. The identification of the bonding within the topmost layer is based on the partial density of states and Crystal Orbital Hamilton Population (COHP) [29,30].

Structural and electronic properties of flat GaN(0001) surface were investigated by *ab initio* simulations using several slab sizes: $(4 \times 4)$, $(6 \times 4)$ and $(8 \times 4)$. The selection was partially determined by the earlier results indicating existence of $(2 \times 1)$ reconstruction [10,11]. The electronic properties, obtained in $(4 \times 4)$ slab are presented in Fig. 1 using the methods previously developed [31]. These data indicate that the surface states bands of GaN (0001) surface are different from the previously published [7-9]. The presently obtained surface states actually belong to two bands, separated by a gap of 0.3 eV. The lower surface state band is fully occupied, the upper is empty thus the Fermi level is located at the top of lower band. The upper surface band (empty) is identified as belonging to the topmost layer $Ga - p_z$ orbitals.

The lower surface band (occupied) consists of the subband that are associated with the top layer $Ga - sp^3$ (upper) and $Ga - sp^2$ (lower) hybridized orbitals. The energy difference leads to emergence of the gap of about 0.3 eV. In summary stoichiometric GaN(0001) surface is semiconducting.

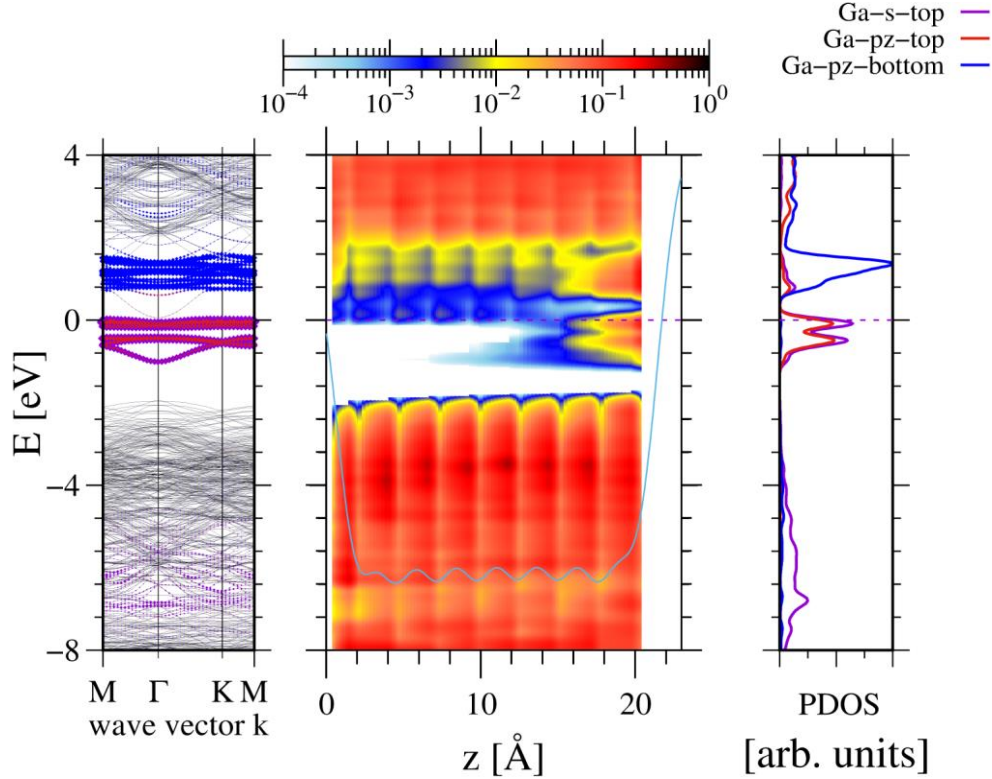

Fig. 1. Electronic properties of (4 × 4) 8 double Ga-N atomic layers (DALs) thick slab, representing stoichiometric Ga-terminated GaN(0001) surface: left – band diagram in momentum space, center - energy bands in real space, plotted along c-axis, right - partial density of states (PDOS) plotted for the top layer Ga atoms. The blue and magenta color represent top layer Ga atom $Ga - sp^3$ hybridized and $Ga - p_z$ orbital related states, respectively.

As it is shown in Fig. 2, these electronic properties are closely related to the structural properties of the surface. The gallium topmost surface layer atoms are divided into two sets: first – located approximately in standard lattice sites and the second - located in the plane of the topmost nitrogen atoms. Therefore, the first set is identified as associated with the $sp^3$ hybridization and the second with $sp^2$ hybridization. In addition, the $Ga-p_z$ state in the second set is higher in energy, located above Fermi level and therefore it is not occupied as shown in Fig 1. The occupied states are different in energy, those of $sp^2$ hybridization have more contribution from $Ga$-$s$ state, therefore they have lower energy. Those of $sp^3$ hybridization have slightly higher energy, thus we observe two peaks in PDOS plot.

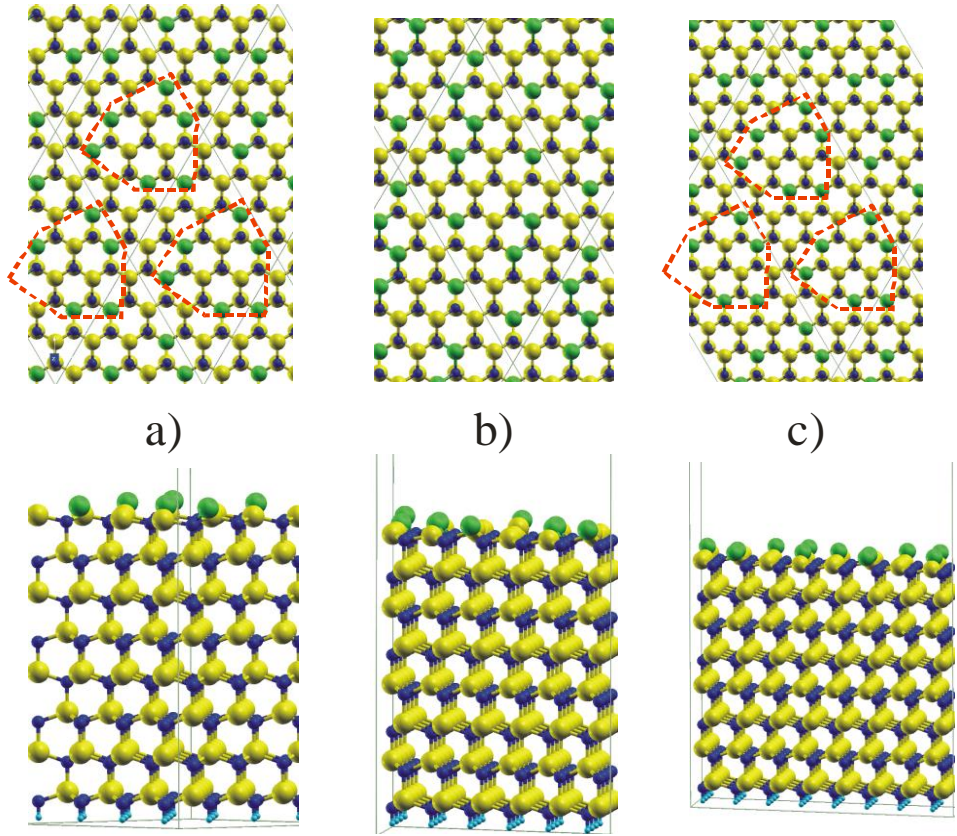

Fig. 2. Gallium nitride 8 DALs thick slabs representing Ga-terminated GaN(0001) surface: a) $(4 \times 4)$ slab, b) $(6 \times 4)$ slab, c) $(8 \times 4)$ slab. The upper and lower rows present the top and side view, respectively. The balls represent the following atoms: blue – nitrogen, yellow – gallium, green – top layer gallium ($sp^2$) hybridized , cyan – hydrogen termination pseudoatoms. The red color broken bond lines indicate the shape of $(4 \times 4)$ reconstruction pattern.

As shown in Fig 2 (a), in the case of $(4 \times 4)$ slab the set of ten ($sp^3$) hybridized Ga atoms is approximately located in the standard GaN lattice sites while the remaining six ($sp^2$) hybridized Ga atoms are positioned in the topmost nitrogen atom plane This could be understood from ECR analysis as follows: denote the fraction of atoms of $sp^2$ hybridization by $x$, therefore the fraction of $sp^3$ hybridization atom is $(1 - x)$. The latter atoms have 2 states that are occupied by the electrons, the former have $p_z$ states empty. Since the average charge for surface Ga top atom broken bond is $(3/4)$, then the electron charge balance equation is:

$$(1-x)2 = \frac{3}{4} \tag{1}$$

From this equation it follows that the fraction of $sp^2$ hybridized Ga atoms is $x = \frac{5}{8}$. The remaining fraction $1 - x = \frac{3}{8}$ Ga atoms is $sp^3$ hybridized. Accordingly, in $(4 \times 4)$ slab 10 Ga atoms should be located in the N atom plane while the remaining 6 Ga atoms should be located in the higher positions. The distribution of Ga atoms in the topmost plane is in full agreement with the ECR prediction. Similarly in the case of $(6 \times 4)$ slab 15 and 9 Ga atoms are in the predicted position in full agreement with the ECR argument. And finally, in the case of $(8 \times 4)$ slab, 20 and 12 Ga atoms are in $sp^2$ and $sp^3$ hybridization, respectively. In summary, the *ab initio* results fully confirms prediction based on ECR analysis, therefore the atoms are selected due to the electron redistribution.

As it is shown in Fig. 1, in the case of $(4 \times 4)$ and $(8 \times 4)$ slabs, the emerging $(4 \times 4)$ reconstruction is compatible with the slab periodicity. In case of $(6 \times 4)$ it is not, therefore this reconstruction is not observed. Thus successful simulations of the surface reconstruction requires proper choice of the simulation slab. Therefore for large size surface $(4 \times 4)$ reconstruction based on electron redistribution is energetically stable configuration of GaN(0001) stoichiometric surface.

An important question arises related to the charge transfer not only between the surface sites but also between surface and the bulk. This was investigated by investigation of p-type and n-type doped slabs. The p-type doping was simulated by introduction of two Mg substitutional atoms in Ga sites that leads to the total deficit of the two electrons in the valence band which in principle should be sufficient to evacuate two electrons from the $sp^3$ hybridized Ga atoms and change the reconstruction of the surface. Similarly, substitution of Ga by Si atoms leads to the surplus of the two electrons valence charge so that these additional charges may be located on the Ga surface bonds, thus increasing number of $sp^3$ hybridized Ga atoms by unity. This could affect also the potential distribution within the slab.

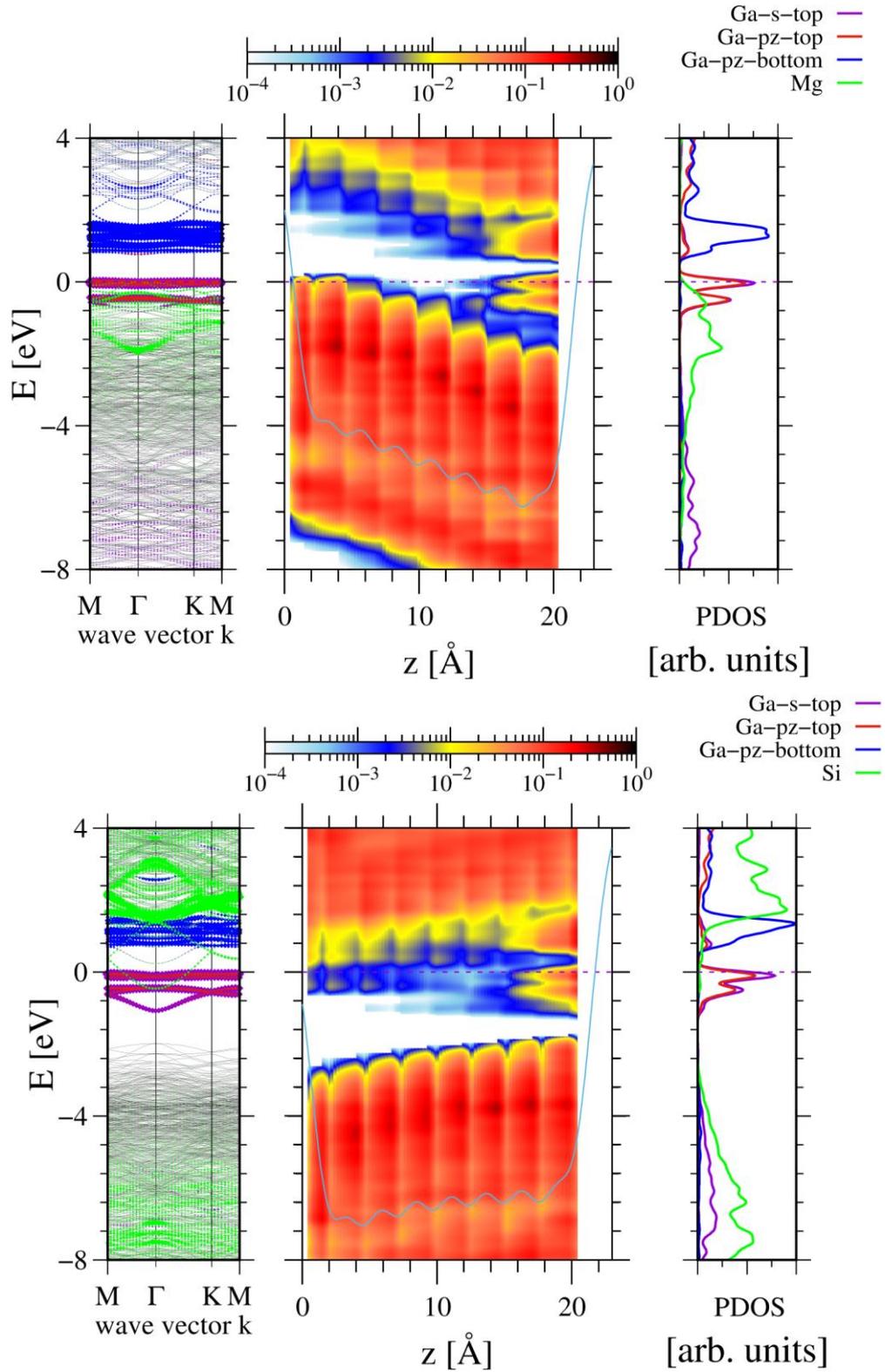

Fig. 3. Electronic properties of (4 × 4) 8 double Ga-N atomic layers (DALs) thick slab, representing Ga-terminated GaN(0001) surface: (a) doped by Ga substitution by two Mg atoms (0.0156 at%), (b) by Ga substitution by two Si atoms (0.0156 at%). The notation is as in Fig 1. Additional green color lines represent Mg and Si states, multiplied by factor of 10 in order to visualize their presence.

As it is shown in Fig. 3, the electric properties of the GaN(0001) surface slabs are considerably affected by the doping, inducing considerable field along c-axis. In case of Mg doping, the electric field estimated from the slope of the potential is $\vec{E}(Mg) \cong 6 \times 10^{-2}\ V/\text{Å} = 6\ MV/cm$. This considerable field related to the positive charge accumulated at the surface that can be assessed from Gauss law $q_{Mg} = \varepsilon_o \epsilon_{GaN} |\vec{E}(Mg)|\ S$. The (4 × 4) slab surface area is $S = 16\ a_{GaN}^2\ \sqrt{3}/2 = 1.406 \times 10^2 \text{Å}^2 = 1.406 \times 10^{-19} m^2$ in which the *ab initio* lattice constant was used i.e. $a = 3.194\ \text{Å}$. The other values used were: vacuum permittivity $\varepsilon_o = 8.854 \times 10^{-12}\ F/m$ and GaN dielectric permittivity $\epsilon_{GaN} = 10.28$. The total charge was therefore $q_{Mg} = 7.68 \times 10^{-20}\ C = 0.48\ e$, i.e. much below the charge $2\ e$ necessary for full occupation of two $sp^3$ hybridized states necessary for transformation of single Ga surface atom. This is confirmed by different occupation of higher $sp^3$ related peak in Fig. 3 (a) and (b). In case of Mg doping (Fig. 3 (a)) the higher peak is occupied in half approximately while in the case of Si doping (Fig. 3 9b)) this occupation is close to complete. Therefore the fractional shift from $sp^2$ to $sp^3$ hybridized configuration was observed. Naturally this is a consequence of the analysis in terms of the hybridized basis set, in fact both types of basic functions are present in all cases, the difference is in their occupation only.

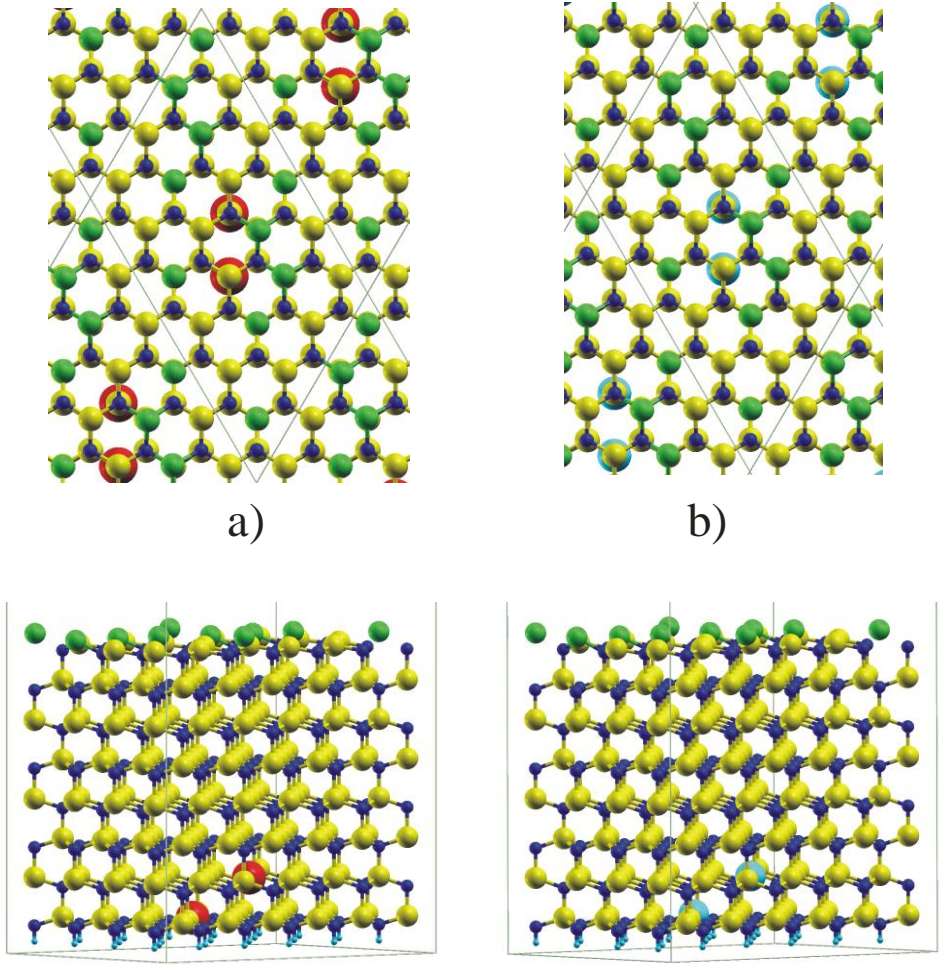

Fig. 4. Gallium nitride $(6 \times 4)$ 8 DALs thick slabs representing Ga-terminated GaN(0001) surface doped with the pair of Si substitutional atoms. The most symbols are as in Fig. 1. Additionally, red and cyan balls represent Mg and Si atoms, respectively.

Similarly, the accumulation of the negative surface charge related to substitution of the two Si substitutional atoms may be assessed. The field related to Si doping is not uniform within the slab, indication important role of the bulk charge. The field at the surface is much lower $\vec{E}(Si) \cong 2 \times 10^{-2}\ V/Å = 2\ MV/cm$, therefore the total charge at the slab surface is $q_{Si} = 2.56 \times 10^{-20}\ C = 0.16\ e$, again not sufficient to induce change of the surface symmetry.

An additional investigation of the doping of $(6 \times 4)$ slab to p-type and n-type was made by substitution of the pair of Ga atoms by Mg and Si respectively. The results, plotted in Fig. 4 indicate that in the case of n-type doping, the number of top Ga atoms in $sp^3$ hybridization was increased to 10, thus finally confirming the decisive role of the charge in GaN(0001) reconstruction. This could be confirmed by the plots of band diagrams in Fig. 5.

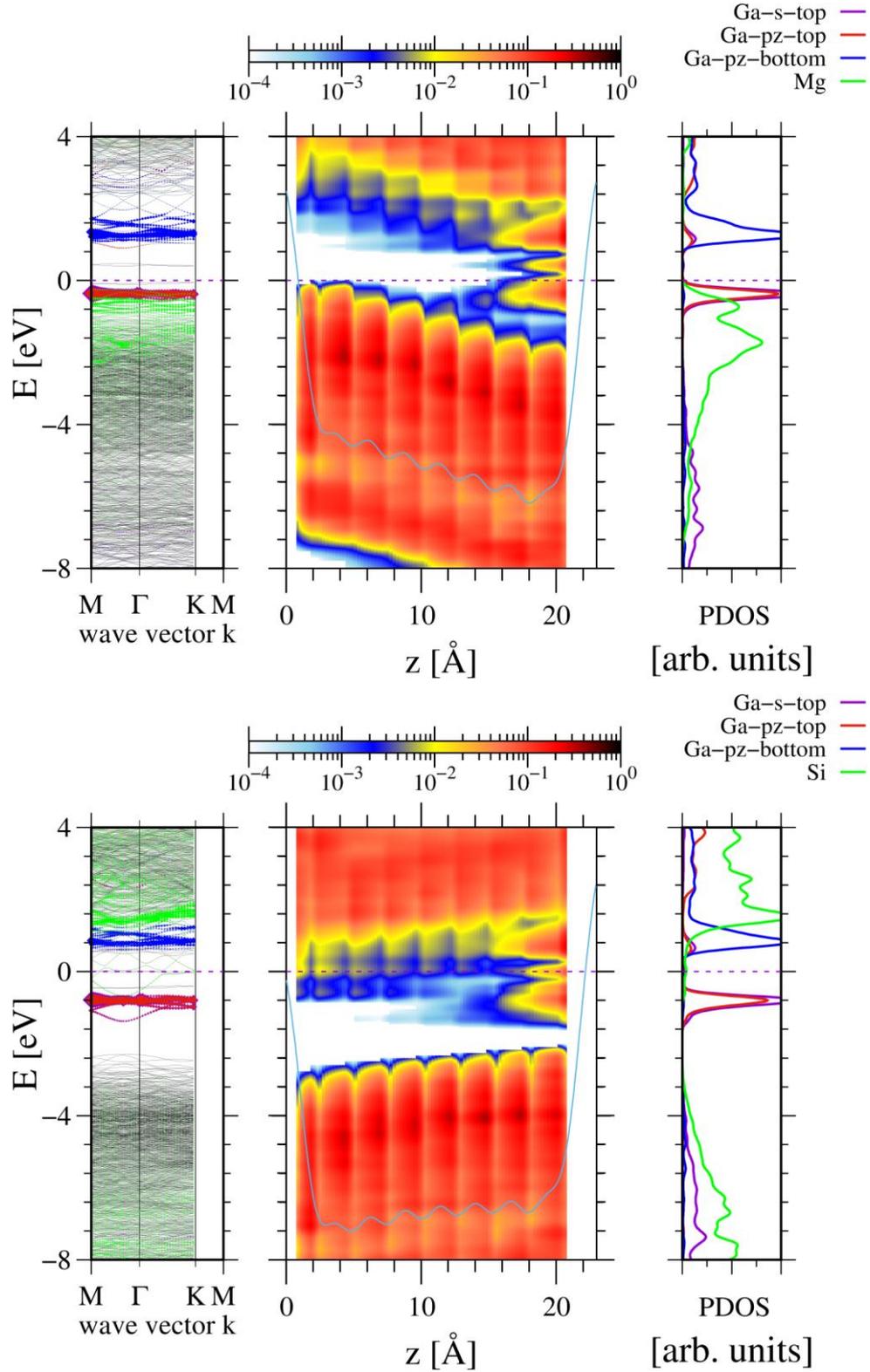

Fig. 5. Electronic properties of (6 × 4) 8 double Ga-N atomic layers (DALs) thick slab, representing Ga-terminated GaN(0001) surface: (a) doped by Ga substitution by two Mg atoms (0.0104 at%), (b) by Ga substitution by two Si atoms (0.0104 at%). The notation is as in Fig 1. Additional green color lines represent Mg and Si states, multiplied by factor of 10 in order to visualize their presence.

In case of Mg doping, the field is $\vec{E}(Mg) \cong 5 \times 10^{-2}\ V/\text{Å} = 5\ MV/cm$, the surface is $S = 24\ a_{GaN}^2\ \sqrt{3}/2 = 2.109 \times 10^2 \text{Å}^2 = 2.109 \times 10^{-19} m^2$ which gives the total charge $q_{Mg} = 1.15 \times 10^{-19}\ C = 0.72\ e$, i.e. still below the critical charge of $2\ e$ thus the fractional change is observed only. In case of Si the numbers are $\vec{E}(Si) \cong 2 \times 10^{-2}\ V/\text{Å} = 2\ MV/cm$, which gives the total charge is $q_{Si} = 3.84 \times 10^{-20}\ C = 0.24\ e$. Nevertheless, as it is shown in Fig 6, the number of $sp^3$ hybridized atoms is increased to 10.

In general, the symmetry of the surface relaxation results from the charge and strain energy optimization. In the case of the slab and reconstruction compatibility, the system is more stable, therefore the change of the reconstruction is more difficult to induce by doping. In case of incompatible $(6 \times 4)$ slab, the number of the differently hybridized atoms could be changed more easily.

The driving force of the above reconstruction is the difference of the energy of $Ga - s$ and $Ga - p$ orbitals. Therefore this reconstruction can occur for surfaces of typical III-V semiconductors such as GaAs(111). In case of the nitrides, Ga terminated surfaces can undergo this reconstruction such as investigated GaN(0001). On the contrary, nitrogen terminated surfaces will not undergo this reconstruction such N-terminated $GaN(000\bar{1})$ surface [27,32]. Despite large difference in the energies of $N - s$ and $N - p$ orbitals, the basic bonding of the solid does not involve $sp^3$ hybridization, therefore such reconstruction is not observed.

The obtained results may be summarized as follows:

i/ charge transfer controls the mixed reconstruction in which $(3 / 8)$ top layer Ga atoms remain in standard position with $sp^3$ hybridized bonding while the remaining $(5/8)$ top layer Ga atoms is located in plane of N atoms with $sp^2$ hybridized bonding,

ii/ the charge controlled hybridization mechanism is different from the bridge type overlap mechanism, frequently observed in reconstructions of semiconductor surfaces,

iii/ $(4 \times 4)$ symmetry results from charge induced reconstruction of Ga-terminated stoichiometric $GaN(0001)$ surface,

iv/ reconstruction induced strain plays important role in stabilizing $(4 \times 4)$ symmetry,

v/ bulk doping affect surface reconstruction symmetry,

vi/ the charge transfer energy reduction mechanism is universal, typical for all surfaces terminated by $sp^3$ bonded atoms.

In summary, these results provide new picture of the properties of stoichiometric semiconductor surfaces.

This research was partially supported by Polish National Science Center grant number DEC- . The calculations reported in this paper were performed using the computing facilities of the Interdisciplinary Centre for Mathematical and Computational Modelling of Warsaw University (ICM UW) under grant GB84-23. We gratefully acknowledge Polish high-performance computing infrastructure PLGrid (HPC Centers: ACK Cyfronet AGH) for providing computer facilities and support within computational grant no. PLG/2023/016668.